\documentclass[letter,twocolumn,aps,showpacs]{revtex4}

\usepackage{graphics}
\usepackage{graphicx}
\usepackage{amssymb}
\usepackage{amsmath}
\usepackage{epsfig}
\usepackage{color}
\usepackage{ulem}
\usepackage{hhline}
\usepackage{multirow}

\begin{document}

\title{Magnetism, Superconductivity and Quantum Criticality in the Multi-Site Cerium Heavy Fermion Compound Ce$_3$PtIn$_{11}$.}

\author{
J.\ Prokle\v{s}ka$^{1}$,
M.\ Kratochv\'{\i}lov\'{a}$^{1}$,
K.\ Uhl\'{\i}\v{r}ov\'{a}$^{1}$,
V.\ Sechovsk\'{y}$^{1}$, and
J.\ Custers$^{1}$
}

\affiliation{$^1$ Charles University in Prague, Faculty of Mathematics and Physics, Ke Karlovu 5, 121 16 Prague 2, Czech
Republic.}


\begin{abstract}
The properties of the novel heavy fermion superconductor Ce$_3$PtIn$_{11}$ are investigated by thermodynamic and transport measurements at ambient and under hydrostatic pressure. At ambient pressure the compound exhibits two successive magnetic transitions at $T_{\mathrm{1}} \simeq 2.2$~K and $T_{\mathrm{N}} \simeq 2$~K into antiferromagnetically ordered states and enters into a heavy fermion superconducting phase below $T_{\mathrm{c}} \simeq 0.32$~K. The coexistence of long-range magnetic order and superconductivity is discussed in the context of the existence of the two crystallographically inequivalent Ce-sites in the unit cell of Ce$_3$PtIn$_{11}$. The experimental data allow us to construct the pressure-temperature phase diagram.
\end{abstract}
\date{\today}

\maketitle

\begin{figure}[t]
\centerline{\includegraphics[width=0.9\columnwidth]{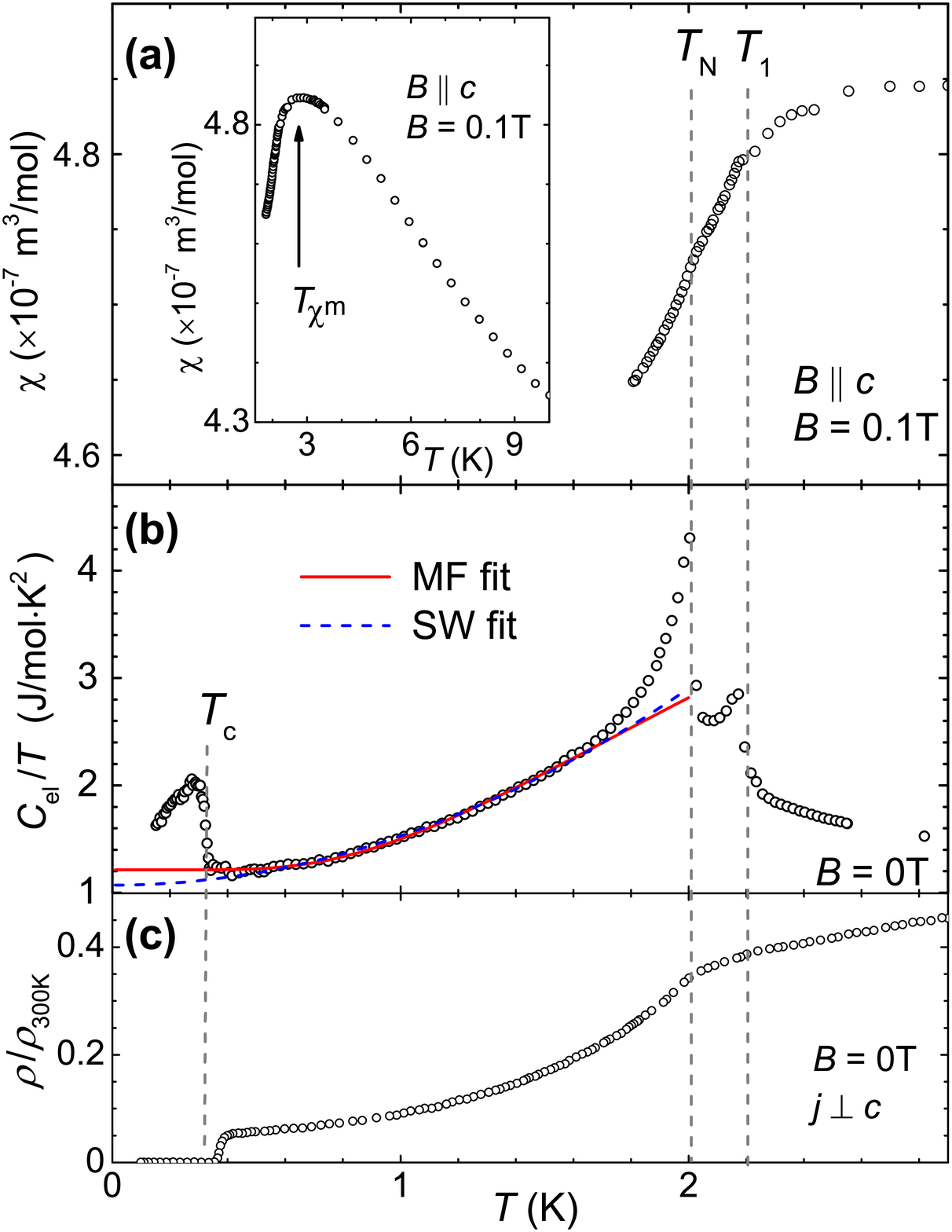}}
\caption{\label{Fig1} Overview of low temperature properties of Ce$_3$PtIn$_{11}$ for $T < 3$~K. (a) Temperature dependence of $\chi$ vs $T$ in field $B = 0.1$~T applied $\parallel c$--axis. Inset shows the data up to 10~K. The arrow indicates the maximum in $\chi(T)$. (b) Electronic part of the specific heat $C_{\mathrm{el}}/T$ as function of $T$ in zero field. The red solid and blue dashed lines are respectively a mean-field (MF) and spin wave (SW) fit of $T < 0.8T_{\mathrm{N}}$ towards $T =0$ (see text). (c) Resistivity normalized to its room temperature value in $B = 0$~T and $j\perp c$. 
}
\end{figure}

\begin{figure}[t]
\centerline{\includegraphics[width=\columnwidth]{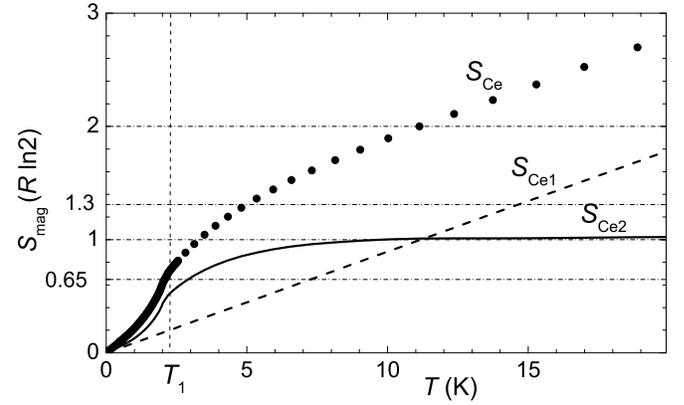}}
\caption{\label{Fig2} The magnetic entropy $S_{\mathrm{mag}}$ associated with $C_{\mathrm{el}}/T$ ($S_{\mathrm{Ce}}$; symbols), the Ce1-sublattice ($S_{\mathrm{Ce1}}$; dashed line) and the Ce2-sublattice ($S_{\mathrm{Ce2}}$; solid line) as discussed in the text.}
\end{figure}

\begin{figure}[t]
\centerline{\includegraphics[width=\columnwidth]{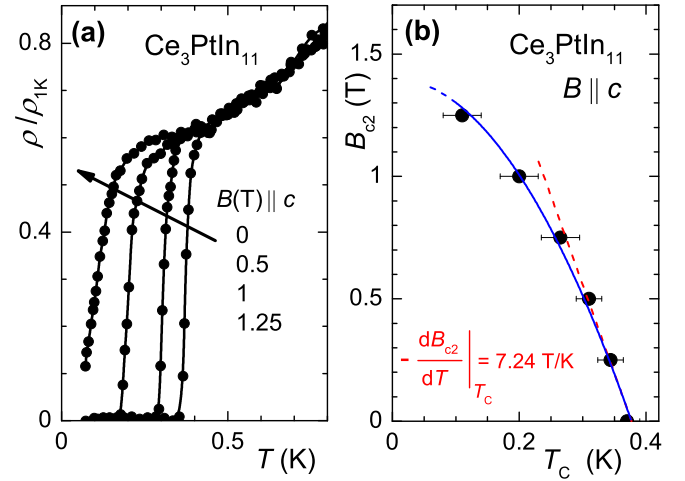}}
\caption{\label{Fig3} Selection of low temperature $\rho(T)$ runs in applied field (a) and the derived phase diagram for $B_{\mathrm{c2}}$ (b). The red dashed line depicts the slope in the vicinity $B_{\mathrm{c2}} \rightarrow 0$~T.}
\end{figure}

\begin{figure}[t]
\centerline{\includegraphics[width=\columnwidth]{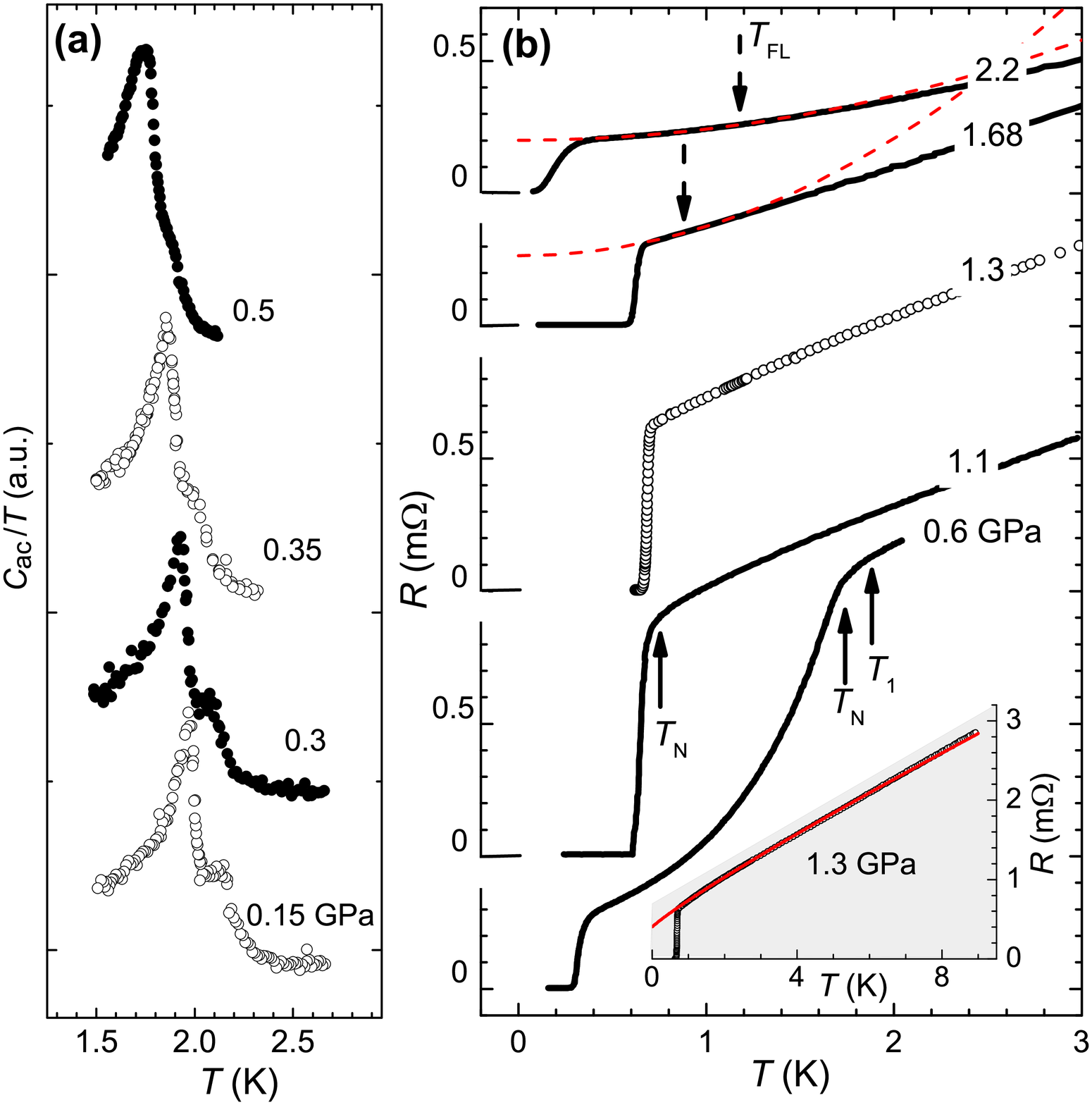}}
\caption{\label{Fig4} Experiments under hydrostatic pressure. (a) Results of ac~calorimetry $C_{\mathrm{ac}}$ plotted in various applied pressures. For better view data has been shifted along $C_{\mathrm{ac}}/T$--axis. (b) Resistance $R$ vs $T$ for some selected pressures. Solid arrows show the magnetic transitions. Dashed arrows mark the upper limit ($T_{\mathrm{FL}}$) of the $R(T) = R_{\mathrm{0}}+AT^2$ fits (red dashed lines). Inset: $R(T)$ at $p_{\mathrm{c}}$ in extended $T$-range with fit (red line).}
\end{figure}

\begin{figure}[t]
\centerline{\includegraphics[width=\columnwidth]{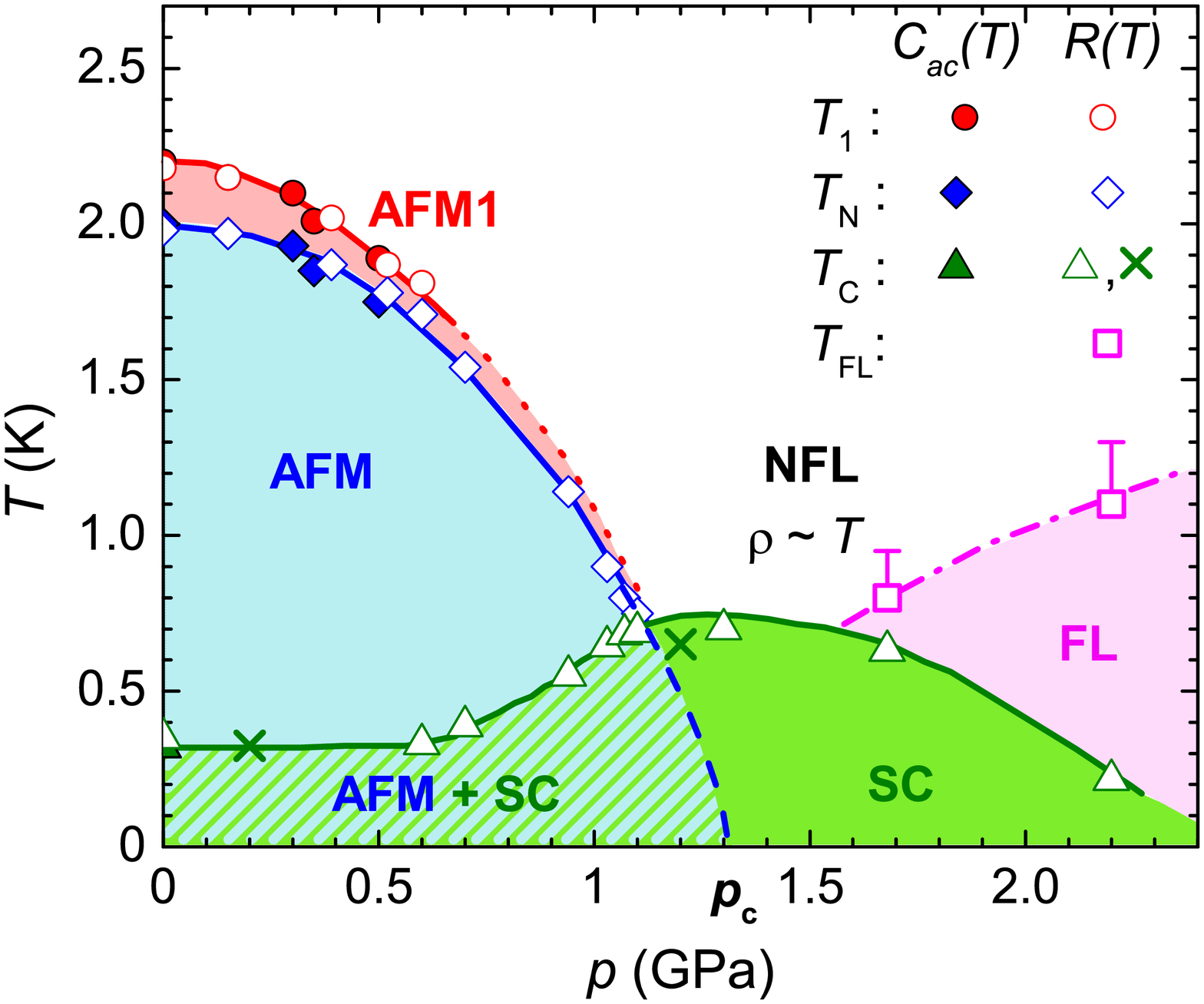}}
\caption{\label{Fig5}  The zero field $p - T$ phase diagram of Ce$_{3}$PtIn$_{11}$. Transition temperatures $T_1$ (circles), $T_{\mathrm{N}}$ (diamonds), $T_{\mathrm{c}}$ (triangles) and $T_{\mathrm{FL}}$ (squares) obtained from $C_{\mathrm{ac}}$ and $R$ are denoted by closed and open symbols, respectively. The crosses are results from a sample from different batch. A possible continuation of $T_{\mathrm{1}} \rightarrow T_{\mathrm{c}}$ ($T_{\mathrm{N}} \rightarrow 0$) is indicated by the red dotted (blue dashed) line. The shaded area marks coexistence of SC and AFM. A purely SC ground state is found for $p > p_{\mathrm{c}}$.}
\end{figure}

In several classes of strongly correlated electron systems such as heavy fermion (HF) materials superconductivity emerges in the vicinity of a quantum critical point (QCP). The general view is that critical fluctuations associated with the order parameter of the phase transition at the critical point enhance the interaction between electrons in a similar way phonons do in conventional superconductors leading to superconductivity (SC)~\cite{Gegenwart2008}. In fact superconductivity has been reported in numerous HF materials to emerge at the border of a magnetically ordered state and a paramagnetic state indicative of magnetically mediated SC~\cite{Monthoux2007,Pfleiderer2009}.\\
The majority of the cerium HF compounds studied to date have one crystallographic site for the cerium ions. In compounds with two or more inequivalent sites, the different local environment of corresponding Ce ions results in different interaction of the Ce 4$f$ states with the surrounding ligand and conduction states. This in turn will lead to different Kondo coupling strength. Consequently, a variety of new and complex phenomena might be expected in these multi--site cerium compounds where the ground state is characterized by a coexistence of different electronic and magnetic states on microscopic scale. Such is well illustrated in cubic Ce$_3$Pd$_{20}$Si$_6$ where one Ce-site exhibits dipolar ($T_{\mathrm{N}}\approx 0.3$~K) and the second site quadrupolar antiferromagnetic (AFM) order ($T_{\mathrm{Q}}\approx 0.5$~K)~\cite{Strydom2006,Goto2009,Mitamura2010}. The compound arouses special attention displaying pronounced non-Fermi liquid behavior and a field-induced QCP ($T_{\mathrm{N}} \rightarrow 0$ for $B_{\mathrm{c}} \approx 0.9$~T) of Kondo breakdown type which separates two different ordered phases~\cite{Custers2012}. Theoretically, little has been investigated on the behavior of multiple distinct Kondo lattices either. Benlagra and co--workers discussed a Kondo lattice comprising two local-moment sublattices, coupled with different Kondo couplings ($T_{\mathrm{K1}}$ and $T_{\mathrm{K2}}$) to conduction electrons, factual describing a compound with two inequivalent, for instance, Ce-sites~\cite{Benlagra2011}. Particularly interesting is the situation of partial screening ($T_{\mathrm{K1}} > T > T_{\mathrm{K2}}$) where one sublattice is in a non-magnetic Kondo screened state forming heavy quasiparticles, while the second sublattice still carries local magnetic moments. Here, any magnetic long--rang ordered (LRO) state, hence, will manifest characteristics of a weakly polarized HF phase coexisting with properties of typical local--moment magnetism. Such has been reported for Ce$_7$Ni$_3$ which possesses even three inequivalent Ce-sites~\cite{Sereni1994}. In this context one may speculate that under certain conditions the HF sublattice becomes superconducting while the second sublattice remains magnetically ordered.\\
Here we introduce a new multi--site cerium HF compound, Ce$_{3}$PtIn$_{11}$. It belongs to the Ce$_{n}T_{m}$In$_{3n+2m}$ class of materials which comprises a numerous amount of compounds including CeCoIn$_5$, CeRhIn$_5$ and Ce$_2$RhIn$_8$~\cite{Thompson2012,Thompson2003}. The crystal structure of Ce$_3$PtIn$_{11}$ is like that of Ce$_{3}$PdIn$_{11}$ (space group $P4/mmm$) when replacing Pd by Pt~\cite{Kratochvilova2014}. The lattice parameters at room temperature are $a = 4.6874(4)$~{\AA} and $c = 16.8422(12)$~{\AA}. The three Ce ions in the Ce$_3$PtIn$_{11}$ unit cell are distributed within two crystallographically inequivalent sites. Two Ce ions reside the Ce1--position (Wyckoff 2$g$ place, $D_{4h}$ symmetry) which are surrounded by ligands similar as the Ce--ions in Ce$_2$PtIn$_8$~\cite{Kratochvilova2014}. The Ce2--site (Wyckoff 1$a$, $C_{4v}$ symmetry) is occupied by 1 Ce ion. The ion experiences CeIn$_{3}$-like environment. At ambient pressure and in the absence of magnetic field the material shows remarkable properties: Ce$_{3}$PtIn$_{11}$ undergoes two successive magnetic transitions at $T_{\mathrm{1}} \simeq 2.2$~K and $T_{\mathrm{N}} \simeq 2$~K into AFM states~\cite{Kratochvilova2014}, and becomes superconducting below $T_{\mathrm{c}} \simeq 0.32$~K (this work). We, thus, consider Ce$_{3}$PtIn$_{11}$ as a promising candidate for {\it (i)} studying the complex behavior of two different mutually interacting Ce-sites and {\it (ii)} examining the interplay of SC and magnetism.\\
Single crystals of Ce$_3$PtIn$_{11}$ used in the present study were grown out of excess In flux. Details about crystal growth and characterization can be found in Ref.~\cite{Kratochvilova2014}. The magnetic susceptibility in the temperature range from 1.8 to 300~K was determined utilizing a commercial MPMS 7~T SQUID magnetometer from Quantum Design (QD) with an applied field of 0.1~T. For the low temperature experiments the sample was mounted to a Leiden Cryogenics MCK72 dilution refrigerator. A maximum field of 9~T can be applied. Measurements to higher temperatures were performed in a QD PPMS 9~T. Electrical resistivity experiments were conducted by standard 4 point ac~technique. The relaxation method was applied to determine the specific heat (temperature range of $0.15 < T < 30$~K). We used a QD dilution refrigerator heat capacity puck for the specific heat measurements in the MCK72. To apply hydrostatic pressure up to about $2.5$~GPa the sample was loaded into a double cylinder CuBe/NiCrAl pressure cell. Daphne 7373 oil was used as pressure medium. All stated pressures are at low temperatures.\\
Before presenting the low temperature properties of Ce$_3$PtIn$_{11}$ we briefly summarize the behavior above 3~K (not shown). The electrical resistivity $\rho$ is similar in magnitude and behavior to the other Ce$_{n}T_{m}$In$_{3n+2m}$ compounds: at 300~K the $c$-axis resistivity equals 60~$\mu\Omega$cm being approximately 1.5 times larger than $\rho$ in basal plane. Upon lowering temperature $\rho$ shows a weak temperature dependence with $d\rho/dT >0$. Below $\approx 30$~K the resistivity drops rapidly marking the crossover from incoherent Kondo scattering at high temperatures to the formation of heavy-electron Bloch states at low temperatures. The susceptibility as a function of temperature, $\chi(T)$, has been measured in magnetic field applied perpendicular ($\perp c$) and along ($\parallel c$) the crystallographic $c$-axis and appears to be weakly anisotropic with a ratio $\chi^{\parallel c}/\chi^{\perp c} \approx 1.25$ at 3~K. Above 150~K $\chi(T)$ follows Curie--Weiss law with value of the effective moment of $\mu_{\mathrm {eff}}=$ 2.60~$\mu_{\mathrm {B}}$/Ce for both directions, in good agreement with expected Hund's rule value for free Ce$^{3+}$-ion (2.54~$\mu_{\mathrm{B}}$). The obtained Weiss temperatures from fitting yield $\theta^{\perp c}_p = -64$~K and $\theta^{\parallel c}_p = -42$~K.\\
Figure~\ref{Fig1} displays the ambient pressure thermodynamic and transport properties of Ce$_{3}$PtIn$_{11}$ at low temperatures. The results are plotted in the same temperature interval (0 - 3~K) one below the other to compare the associated anomalies and to complement the methods among each other. The inset in Fig.~\ref{Fig1}a shows the low-$T$ susceptibility $\chi^{\parallel c}$ in a wider temperature range. A broad structure with a maximum at $T_{\chi^{\mathrm{m}}} = 2.8$~K is seen which we attribute to the presence of short-range AFM correlations~\cite{Fisher1962}. At slightly lower temperatures these correlations lead to AFM ordering (AFM1) manifested by a sharp decline in $\chi(T)$ at $T_{\mathrm{1}}$ (Fig.~\ref{Fig1}a). The second AFM transition at $T_{\mathrm{N}} \simeq 2.0$~K appears as a weak bump-like structure in the 0.1~T data. The zero field electronic part of the specific heat $C_{\mathrm{el}}/T$ displayed in Fig.~\ref{Fig1}b has been determined by subtracting the lattice term $\beta$ of the Debye fit ($C/T = \gamma_{\mathrm{0}} + \beta T^2$ ; fit interval $10< T <30$~K) from the total specific heat, i.\,e.\,, $C_{\mathrm{el}}/T = C/T - \beta T^2$. The fit yields a Sommerfeld coefficient $\gamma_{0} = 0.52$~J/(mol~K$^{2}$) and $\beta = 5.58$~mJ/(mol~K$^{4}$). The latter value transforms into a Debye temperature of $\Theta_{\mathrm{D}} = 174$~K. The contribution arising from the crystal electric field can be neglected for these low temperatures. The magnetic transitions are clearly visible as pronounced jumps in $C_{\mathrm{el}}/T$. The higher temperature transition is characterized by a shallow kink in the temperature dependence of the electrical resistivity only. In contrast, the transition at the lower temperature is manifested as a steep decrease of $\rho$. It strongly mimics the behavior of $\rho(T)$ for CeRhIn$_5$ for $T<T_{\mathrm{N}} = 3.8$~K~\cite{Hegger2000}. We define the midpoint of the $C_{\mathrm{el}}/T$ jumps as transition temperatures yielding $T_{\mathrm{1}} \simeq 2.2$~K and $T_{\mathrm{N}} \simeq 2$~K, respectively. These values coincide reasonably with the features in $\chi(T)$ and $\rho(T)$ (dashed lines in Fig.~\ref{Fig1}). The most intriguing feature in $C_{\mathrm{el}}/T$ is observed around 0.35~K. It signals the transition into a superconducting phase as corroborated from resistivity data. The jump is slightly broadened and hence we define the transition temperature at the midpoint of the jump height becoming $T_{\mathrm{c}} \simeq 0.32$~K. To estimate the normal state specific heat coefficient $\gamma_{\mathrm{n}}$ we extrapolated the low-$T$ tail of the AFM transition towards $T =0$ (red dashed line in Fig.~\ref{Fig1}b) employing a second-order mean-field type expression, $C_{\mathrm{el}}/T = \gamma_{n} + A\exp^{-\Delta_{\mathrm{AFM}}/T}$~\cite{Maple1986}. The resulting parameters are $\gamma_{\mathrm{n}} = 1.21$~J/(mol~K$^2$), $A = 8.97$~J/(mol~K$^2$) and $\Delta_{\mathrm{AFM}} = 3.44$~K (fit interval $0.44 <T <1.6$~K). As shown, a fit $C_{\mathrm{el}}/T \propto T^2$ (AFM spin wave\cite{Kranendonk1958}) describes the data in the same interval with similar quality but with a reduction in $\gamma_{\mathrm{n}}$ by $\approx 10$~\%. Using above parameters allows calculating the parameter $\Delta C/(\gamma_{\mathrm{n}}T_{\mathrm{c}}) \approx 0.7$, that is roughly half of the expected value from BCS theory. However, we implicitly assumed here that the electrons in the conduction band participating in SC originate in equal measures from the two Ce-sites. This assumption might not be correct as discussed later on. The magnetic entropy, $S_{\mathrm{mag}} = \int_{0}^{T} C_{\mathrm{el}}/T dT$, with the mean-field expression replacing the SC part is presented in Fig.~\ref{Fig2}.\\
To provide additional information about the superconducting state we determined the upper critical field $B_{\mathrm{c2}}$ using data depicted in Fig.~\ref{Fig3}. The midpoint of the jump defines $T_{\mathrm{c}}$, which is shown in Fig.~\ref{Fig3}b. The expression for orbital limited superconductivity, $B_{\mathrm{c2}} = B_{\mathrm{c2}}^{\mathrm{orb}}(T=0)[1-(T/T_{\mathrm{c}})^2]$, describes the data reasonably well with $B_{\mathrm{c2}}^{\mathrm{orb}}(0) = 1.4$~T and an initial slope $-dB_{\mathrm{c2}}/dT|_{T=Tc} = 7.24$~T/K. Because $B_{\mathrm{c2}}^{\mathrm{orb}} \propto (m^\ast)^2T_{\mathrm{c}}^2$ it is evident that heavy quasiparticles, mass $m^\ast$, are involved in Cooper-pairing.\\
In the following we investigate the influence of hydrostatic pressure on the magnetic and SC transitions by means of ac calorimetry and resistance measurements (Fig.~\ref{Fig4}a and b). The $p-T$ phase diagram in Fig.~\ref{Fig5} collects the pressure results. We observe that the magnetic transition temperatures decrease with increasing pressure and become absent once they intersect with superconductivity at $p \simeq 1.1$~GPa. For pressures between 1.1 and 1.6~GPa superconductivity evolves out of a non-Fermi liquid state which is characterized by an almost $T$-linear resistivity from $T_{\mathrm{c}}$ ($\simeq 0.7$~K) to temperatures even higher than $T >5$~K. The maximum in $T_{\mathrm{c}}$ points to position of the magnetic QCP being located at $\approx 1.3$~GPa~\cite{Monthoux2007,Monthoux1999}. A fit to $R(T)$ at $p_{\mathrm{c}}$ gives $R = R_0 + AT^n$ with $n=0.90 \pm 0.05$ for $0.8 \leq T \leq 7$~K (inset Fig.~\ref{Fig4}b). A similar exponent has also been reported above $T_{\mathrm{c}}$ for CeRhIn$_5$~\cite{Muramatsu2001} and Ce$_2$RhIn$_8$~\cite{Nicklas2003} near their critical pressures raises speculation that also in Ce$_3$PtIn$_{11}$ the quantum criticality is local. However, within our present data an AFM spin-density wave type of QCP which predicts $R(T) \propto T^{d/z}$, where the dynamical exponent $z = 2$ and $d$ is the effective dimensionality of the critical spin-fluctuation~\cite{Millis1993} cannot be excluded. For pressures higher than $p > 1.68$~GPa we find that SC emerges out of a Fermi-liquid (FL) phase.\\
A possible scenario for understanding  the properties of Ce$_{3}$PtIn$_{11}$  can be preformed when taking some specific features of the crystal structure into consideration (3 Ce atoms/f.u. : 2Ce1 + 1Ce2). Because of its CeIn$_3$-like environment of the Ce2-site one might assume that the Ce2-sublattice orders AFM while the Ce1-sublattice, which is characterized by an ``Ce$_2$PtIn$_8$'' surrounding as being fully Kondo compensated. Consequently, the heavy QPs associated with the Ce1-sublattice take part in the SC condensate. Supports for this scenario might be found in thermodynamics. In the following we perform a crude analysis of our specific heat data based on two independent Ce-sublattices despite the fact that this is not justified by the theory~\cite{Benlagra2011}. We postulate that $\gamma_{\mathrm{0}} = 0.52$~J/(mol~K$^2$) from the Debye fit is associated with Ce1-sublattice and subtract this term from the total electronic specific heat to obtain the Ce2-sublattice contribution, i.\,e.\,, $C_{\mathrm{Ce2}}/T = C_{\mathrm{el}}/T - \gamma_{0}$. Figure~\ref{Fig2} displays the corresponding entropies of both sublattices. As can be seen, the entropy of the Ce2-sublattice, $S_{\mathrm{Ce2}}$, liberated at $T_{1}$ is found to be $0.5R\ln2$ indicating that the ordered moment is lowered by Kondo interactions. The corresponding sublattice Kondo temperature is $T_{\mathrm{K2}} \simeq 3$~K ($S_{\mathrm{Ce2}} = 0.65 R\ln2$)~\cite{Desgranges1982}. The large $S_{\mathrm{Ce2}}$ below $T_{\mathrm{1}}$ as well as the fact that $T_{\mathrm{K2}}$ is of the same order of magnitude as $T_{\mathrm{1}}$ would suggest the presence of local moments. According to our postulate, $C_{\mathrm{Ce1}}/T$ gives a constant contribution over the entire $T$-range ($0 <T <30$~K). Such is unphysical, however the constant temperature dependence might approximate the HF behavior in the low-$T$ region well. The associated entropy of the Ce1-sublattice, $S_{\mathrm{Ce1}}$, is depicted in Fig.~\ref{Fig2}. $T_{\mathrm{K1}}$ equals 15~K which is of the same order as found in Ce$_2$PdIn$_8$ ($T_{\mathrm{K}} =10$~K)~\cite{Kaczorowski2009}. Recognizing the assumption that the Ce1-sublattice is responsible for SC would imply that the normal state specific heat coefficient is not $\gamma_{\mathrm{n}}$ as it contains contribution from the Ce2-sublattice, but equals $\gamma_{\mathrm{0}}$. Hence one obtains a value of $\Delta C/(\gamma_{\mathrm{n}} T_{\mathrm{c}}) \approx 1.6$, in fair agreement with the BCS value. This would support the coexistence of AFM and SC in a large part of the $p-T$ phase diagram (see Fig.~\ref{Fig5}) as a consequence of different ground state of each sublattice.\\
In summary, Ce$_3$PtIn$_{11}$ is a heavy fermion compound exhibiting two magnetic transitions at $T_{\mathrm{1}} \simeq 2.2$~K and $T_{\mathrm{N}} \simeq 2.0$~K and a superconducting one at $T_{\mathrm{c}} \simeq 0.32$~K at ambient pressure. Both, the AFM order and SC coexist in a large region of the $p-T$ phase diagram up to $p_{\mathrm{c}} \simeq 1.3$~GPa, the compound's magnetic QCP. While still speculative, the QCP is of local-moment type. We suggest that the observed unusual properties can be related to the fact that Ce$_{3}$PtIn$_{11}$ possesses two inequivalent Ce-sites with distinct different Kondo scales. However, the present state of knowledge of this new and interesting compound allowed us to perform only simple data analysis in terms of two independent Ce sublattices. Microscopic experiments (neutron scattering, NQR and NMR) are highly desirable to obtain relevant information on the coupling mechanism between the two Ce sublattices and thereby provide the basis for the theoretical description.  \\

We would like to thank M.M. Abd-Elmeguid for critical reading of the manuscript and J.G. Sereni for helpful discussions. This work was supported by the Grant Agency of the Czech Science Foundation (Project P203/12/1201). Experiments were performed in MLTL (http://mltl.eu/) which is supported within the program of Czech Research Infrastructures (project no. LM2011025).


\vspace{0.5cm}

\bibliographystyle{apsrev4-1}

\end{document}